\documentclass[12pt]{article}
\begin{document}
\begin{center}
{\large\bf Scalar-Tensor-Vector Gravity Theory} \vskip 0.3 true in
{\large J. W. Moffat} \vskip 0.3 true in {\it The Perimeter
Institute for Theoretical Physics, Waterloo, Ontario, N2J 2W9,
Canada} \vskip 0.3 true in and \vskip 0.3 true in {\it Department
of Physics, University of Waterloo, Waterloo, Ontario N2Y 2L5,
Canada}
\end{center}
\begin{abstract}%
A covariant scalar-tensor-vector gravity theory is developed which
allows the gravitational constant $G$, a vector field coupling
$\omega$ and the vector field mass $\mu$ to vary with space and
time. The equations of motion for a test particle lead to a
modified gravitational acceleration law that can fit galaxy
rotation curves and cluster data without non-baryonic dark matter.
The theory is consistent with solar system observational tests.
The linear evolutions of the metric, vector field and scalar field
perturbations and their consequences for the observations of the
cosmic microwave background are investigated.

\end{abstract}
\vskip 0.2 true in e-mail: john.moffat@utoronto.ca


\section{Introduction}

Two theories of gravity called the nonsymmetric gravity theory
(NGT)~\cite{Moffat} and the metric-skew-tensor gravity (MSTG)
theory~\cite{Moffat2} have been proposed to explain the rotational
velocity curves of galaxies, clusters of galaxies and cosmology
without dark matter. A fitting routine for galaxy rotation curves
has been used to fit a large number of galaxy rotational velocity
curve data, including low surface brightness (LSB), high surface
brightness (HSB) and dwarf galaxies with both photometric data and
a two-parameter core model without non-baryonic dark
matter~\cite{Moffat2,Brownstein}. The fits to the data are
remarkably good and for the photometric data only the one
parameter, the mass-to-light ratio $\langle M/L\rangle$, is used
for the fitting, once two parameters $M_0$ and $r_0$ are
universally fixed for galaxies and dwarf galaxies. The fits are
close to those obtained from Milgrom's MOND acceleration law in
all cases considered~\cite{Milgrom}. A large sample of X-ray mass
profile cluster data has also been fitted~\cite{Brownstein2}.

The gravity theories require that Newton's constant G, the
coupling constant $\gamma_c$ that measures the strength of the
coupling of the skew field to matter and the mass $\mu$ of the
skew field, vary with distance and time, so that agreement with
the solar system and the binary pulsar PSR 1913+16 data can be
achieved, as well as fits to galaxy rotation curve data and galaxy
cluster data. In ref.~\cite{Moffat2}, the variation of these
constants was based on a renormalization group (RG) flow
description of quantum gravity theory formulated in terms of an
effective classical action~\cite{Reuter}. Large infrared
renormalization effects can cause the effective $G$, $\gamma_c$,
$\mu$ and the cosmological constant $\Lambda$ to run with momentum
$k$ and a cutoff procedure leads to a space and time varying $G$,
$\gamma_c$ and $\mu$, where $\mu=1/r_0$ and $r_0$ is the effective
range of the skew symmetric field.

In the following, we shall pursue an alternative relativistic
gravity theory based on scalar-tensor-vector gravity (STVG), in
which $G$, a vector field coupling constant $\omega$ and the mass
$\mu$ of the vector field are dynamical scalar fields that allow
for an {\it effective} description of the variation of these
``constants'' with space and time. We shall not presently consider
the variation of the cosmological constant $\Lambda$ with space
and time.

The gravity theory leads to the same modified acceleration law
obtained from NGT and MSTG for weak gravitational fields and the
same fits to galaxy rotation curve and galaxy cluster data, as
well as to agreement with the solar system and pulsar PSR 1913+16
observations.  An important constraint on gravity theories is the
bounds obtained from weak equivalence principle tests and the
existence of a ``fifth'' force, due to the exchange of a massive
vector boson~\cite{Adelberger}. These bounds are only useful for
distances $\leq 100\,A.U.$ and they cannot rule out gravity
theories that violate the weak equivalence principle or contain a
fifth force at galactic and cosmological distance scales. Since
the variation of $G$ in our modified gravity theory leads to
consistency with solar system data, then we can explore the
consequences of our STVG theory without violating any known local
observational constraints.

An important feature of the NGT, MSTG and STVG theories is that
the modified acceleration law for weak gravitational fields has a
{\it repulsive} Yukawa force added to the Newtonian acceleration
law. This corresponds to the exchange of a massive spin 1 boson,
whose effective mass and coupling to matter can vary with distance
scale. A scalar component added to the Newtonian force law would
correspond to an {\it attractive} Yukawa force and the exchange of
a spin 0 particle. The latter acceleration law cannot lead to a
satisfactory fit to galaxy rotation curves and galaxy cluster
data.

In Section 8, we investigate a cosmological solution based on a
homogeneous and isotropic Friedmann-Lema\^{i}tre-Robertson-Walker
(FLRW) spacetime. We present a solution that can possibly fit the
acoustic peaks in the CMB power spectrum by avoiding significant
suppression of the baryon perturbations~\cite{Silk}, and which can
possibly be made to fit the recent combined satellite data for the
power spectrum without non-baryonic dark matter.

All the current applications of the three gravity theories that
can be directly confronted with experiment are based on weak
gravitational fields. To distinguish the theories, it will be
necessary to obtain experimental data for strong gravitational
fields e.g. black holes. Moreover, confronting the theories with
cosmological data may also allow a falsification of the gravity
theories. Recently, the NGT and MSTG were studied to derive
quantum fluctuations in the early universe from an
inflationary-type scenario~\cite{Prokopec}.

\section{Action and Field Equations}

Our action takes the form
\begin{equation}
S=S_{\rm Grav}+S_\phi+S_S+S_M,
\end{equation}
where
\begin{equation}
S_{\rm Grav}=\frac{1}{16\pi}\int d^4x\sqrt{-g}\biggl[
\frac{1}{G}(R+2\Lambda)\biggr],
\end{equation}
\begin{equation}
S_\phi=-\int
d^4x\sqrt{-g}\biggl[\omega\biggl(\frac{1}{4}B^{\mu\nu}B_{\mu\nu} +
V(\phi)\biggr)\biggr],
\end{equation}
and
\begin{equation}
\label{Saction} S_S=\int
d^4x\sqrt{-g}\biggl[\frac{1}{G^3}\biggl(\frac{1}{2}g^{\mu\nu}\nabla_\mu
G\nabla_\nu G-V(G)\biggr)
$$ $$
+\frac{1}{G}\biggl(\frac{1}{2}g^{\mu\nu}\nabla_\mu\omega
\nabla_\nu\omega-V(\omega)\biggr)
+\frac{1}{\mu^2G}\biggl(\frac{1}{2}g^{\mu\nu}\nabla_\mu\mu\nabla_\nu\mu
-V(\mu)\biggr)\biggr].
\end{equation}
Here, we have chosen units with $c=1$, $\nabla_\mu$ denotes the
covariant derivative with respect to the metric $g_{\mu\nu}$. We
adopt the metric signature $\eta_{\mu\nu}={\rm diag}(1,-1,-1,-1)$
where $\eta_{\mu\nu}$ is the Minkowski spacetime metric. We have
\begin{equation}
R_{\mu\nu}=\partial_\lambda{\Gamma^\lambda}_{\mu\nu}
-\partial_\nu{\Gamma^\lambda}_{\mu\lambda}
+{\Gamma^\lambda}_{\mu\nu}{\Gamma^\sigma}_{\lambda\sigma}
-{\Gamma^\sigma}_{\mu\lambda}{\Gamma^\lambda}_{\nu\sigma},
\end{equation}
where $\Gamma^\lambda_{\mu\nu}$ denotes the Christoffel
connection:
\begin{equation}
\Gamma^\lambda_{\mu\nu}=\frac{1}{2}g^{\lambda\sigma}(\partial_\nu
g_{\mu\sigma}+\partial_\mu g_{\nu\sigma}-\partial_\sigma
g_{\mu\nu}),
\end{equation}
and $R=g^{\mu\nu}R_{\mu\nu}$. Moreover, $V(\phi)$ denotes a
potential for the vector field $\phi^\mu$, while $V(G), V(\omega)$
and $V(\mu)$ denote the three potentials associated with the three
scalar fields $G(x),\omega(x)$ and $\mu(x)$, respectively. The
field $\omega(x)$ is dimensionless and $\Lambda$ denotes the
cosmological constant. Moreover,
\begin{equation}
B_{\mu\nu}=\partial_\mu\phi_\nu-\partial_\nu\phi_\mu.
\end{equation}

The total energy-momentum tensor is given by
\begin{equation}
T_{\mu\nu}=T_{M\mu\nu}+T_{\phi\mu\nu}+T_{S\mu\nu},
\end{equation}
where $T_{M\mu\nu}$ and $T_{\phi\mu\nu}$ denote the ordinary
matter energy-momentum tensor and the energy-momentum tensor
contribution of the $\phi_\mu$ field, respectively, while
$T_{S\mu\nu}$ denotes the scalar $G$, $\omega$ and $\mu$
contributions to the energy-momentum tensor. We have
\begin{equation}
\frac{2}{\sqrt{-g}}\frac{\delta S_M}{\delta
g^{\mu\nu}}=-T_{M\mu\nu},\quad\frac{2}{\sqrt{-g}}\frac{\delta
S_\phi}{\delta g^{\mu\nu}} =-T_{\phi\mu\nu},\quad
\frac{2}{\sqrt{-g}}\frac{\delta S_S}{\delta
g^{\mu\nu}}=-T_{S\mu\nu}.
\end{equation}
The matter current density $J^\mu$ is defined in terms of the
matter action $S_M$:
\begin{equation}
\frac{1}{\sqrt{-g}}\frac{\delta S_M}{\delta\phi_\mu}=-J^\mu.
\end{equation}

We obtain from the variation of $g^{\mu\nu}$, the field equations
\begin{equation}
\label{Einsteineqs}G_{\mu\nu}-g_{\mu\nu}\Lambda+Q_{\mu\nu}=8\pi
GT_{\mu\nu},
\end{equation}
where $G_{\mu\nu}=R_{\mu\nu}-\frac{1}{2}g_{\mu\nu}R$. We have
\begin{equation}
Q_{\mu\nu}=G(\nabla^\alpha\nabla_\alpha\Theta
g_{\mu\nu}-\nabla_\mu\nabla_\nu\Theta),
\end{equation}
where $\Theta(x)=1/G(x)$. The quantity $Q_{\mu\nu}$ results from a
boundary contribution arising from the presence of second
derivatives of the metric tensor in $R$ in $S_{\rm Grav}$. These
boundary contributions are equivalent to those that occur in
Brans-Dicke gravity theory~\cite{Dicke}. We also have
\begin{equation}
T_{\phi\mu\nu}=\omega\biggl[{B_\mu}^\alpha B_{\nu\alpha}
-g_{\mu\nu}\biggl(\frac{1}{4}B^{\rho\sigma}B_{\rho\sigma}
+V(\phi)\biggr)+2\frac{\partial V(\phi)}{\partial
g^{\mu\nu}}\biggr].
\end{equation}
The $G(x)$ field yields the energy-momentum tensor:
\begin{equation}
T_{G\mu\nu}=-\frac{1}{G^3}\biggl[\nabla_\mu G\nabla_\nu
G-2\frac{\partial V(G)}{\partial g^{\mu\nu}}
-g_{\mu\nu}\biggl(\frac{1}{2}\nabla_\alpha G \nabla^\alpha
G-V(G)\biggr)\biggr].
\end{equation}
Similar expressions can be obtained for $T_{\omega\,\mu\nu}$ and
$T_{\mu\,\mu\nu}$.

From the Bianchi identities
\begin{equation}
\nabla_\nu G^{\mu\nu}=0,
\end{equation}
and from the field equations (\ref{Einsteineqs}), we obtain
\begin{equation}
\label{conservation} \nabla_\nu T^{\mu\nu}+\frac{1}{G}\nabla_\nu
GT^{\mu\nu}-\frac{1}{8\pi G}\nabla_\nu Q^{\mu\nu}=0.
\end{equation}

A variation with respect to $\phi_\mu$ yields the equations
\begin{equation}
\label{phifieldequation}\nabla_\nu B^{\mu\nu}+\frac{\partial
V(\phi)}{\partial\phi_\mu}+\frac{1}{\omega}\nabla_\nu\omega
B^{\mu\nu}=-\frac{1}{\omega}J^\mu.
\end{equation}
Taking the divergence of both sides with respect to $\nabla_\mu$,
we get
\begin{equation}
\nabla_\mu\biggl(\frac{\partial V(\phi)}{\partial
\phi_\mu}\biggr)+\nabla_\mu\biggl(\frac{1}{\omega}\nabla_\nu
\omega B^{\mu\nu}\biggr)
=-\nabla_\mu\biggl(\frac{1}{\omega}J^\mu\biggr).
\end{equation}
If we assume that the current density $J^\mu$ is conserved, we
obtain the equation
\begin{equation}
\nabla_\mu\biggl(\frac{\partial V(\phi)}{\partial
\phi_\mu}\biggr)+\nabla_\mu\biggl(\frac{1}{\omega}\nabla_\nu
\omega B^{\mu\nu}\biggr)
=-\nabla_\mu\biggl(\frac{1}{\omega}\biggr)J^\mu.
\end{equation}
In standard Maxwell-Proca theory, the conservation of the current
$J^\mu$ is a separate physical assumption.

We shall choose for the potential $V(\phi)$:
\begin{equation}
\label{Vpotential}
V(\phi)=-\frac{1}{2}\mu^2\phi^\mu\phi_\mu+W(\phi),
\end{equation}
where $W(\phi)$ denotes a vector field $\phi^\mu$ self-interaction
contribution. We can choose as a model for the self-interaction:
\begin{equation}
\label{Wself} W(\phi)=\frac{1}{4}g(\phi^\mu\phi_\mu)^2,
\end{equation}
where $g$ is a coupling constant.

The effective gravitational ``constant'' $G(x)$ satisfies the
field equations
\begin{equation}
\label{Gequation} \nabla_\alpha\nabla^\alpha
G+V'(G)+N=\frac{1}{2}G^2\biggl(T+\frac{\Lambda}{4\pi G}\biggr),
\end{equation}
where
\begin{equation}
N=\frac{3}{G}\biggl(\frac{1}{2}\nabla_\alpha G\nabla^\alpha G
-V(G)\biggr)+G\biggl(\frac{1}{2}\nabla_\alpha\omega\nabla^\alpha\omega
-V(\omega)\biggr)
$$ $$
-3\Theta\nabla_\alpha G \nabla^\alpha G
+\frac{G}{\mu^2}\biggl(\frac{1}{2}\nabla_\alpha
\mu\nabla^\alpha\mu-V(\mu)\biggr)+\frac{3G^2}{16\pi}\nabla_\alpha\nabla^\alpha\Theta,
\end{equation}
and $T=g^{\mu\nu}T_{\mu\nu}$. The scalar field $\omega(x)$ obeys
the field equations
\begin{equation}
\label{omegaequation} \nabla_\nu\nabla^\nu\omega+V'(\omega)+F=0,
\end{equation}
where
\begin{equation}
F=-\Theta\nabla_\alpha G\nabla^\alpha\omega
+G\biggl(\frac{1}{4}B^{\mu\nu}B_{\mu\nu}+V(\phi)\biggr).
\end{equation}
The field $\mu(x)$ satisfies the equations
\begin{equation}
\label{muequation} \nabla_\alpha\nabla^\alpha\mu+V'(\mu)+P=0,
\end{equation}
where
\begin{equation}
P=-\biggl[\Theta\nabla^\alpha G
\nabla_\alpha\mu+\frac{2}{\mu}\nabla^\alpha\mu\nabla_\alpha\mu
+\omega\mu^2 G\frac{\partial V(\phi)}{\partial\mu}\biggr],
\end{equation}
and the last term arises from the $\mu$ dependence of $V(\phi)$ in
(\ref{Vpotential}).

If we adopt the condition
\begin{equation}
\nabla^\nu\phi_\nu=0,
\end{equation}
then (\ref{phifieldequation}) takes the form
\begin{equation}
\label{phiboxequation}
\nabla^\nu\nabla_\nu\phi_\mu-{R_\mu}^\nu\phi_\nu+\mu^2\phi_\mu-\frac{\partial
W(\phi)}{\partial\phi^\mu}-\frac{1}{\omega}\nabla^\nu\omega
B_{\mu\nu}=\frac{1}{\omega}J_\mu.
\end{equation}

The test particle action is given by
\begin{equation}
S_{TP}=-m\int d\tau-\lambda\int
d\tau\omega\phi_\mu\frac{dx^\mu}{d\tau},
\end{equation}
where $\tau$ is the proper time along the world line of the test
particle and $m$ and $\lambda$ denote the test particle mass and
coupling constant, respectively. The stationarity condition
$\delta S_{TP}/\delta x^\mu=0$ yields the equations of motion for
the test particle
\begin{equation}
\label{eqsmotion} m\biggl(\frac{d^2x^\mu}{d\tau^2}
+\Gamma^\mu_{\alpha\beta}\frac{dx^\alpha}{d\tau}\frac{dx^\beta}{d\tau}\biggr)
=f^\mu,
\end{equation}
where
\begin{equation}
\label{force} f^\mu=\lambda\omega {B^\mu}_\nu\frac{dx^\nu}{d\tau}
+\lambda\nabla^\mu\omega
\biggl(\phi_\alpha\frac{dx^\alpha}{d\tau}\biggr)
-\lambda\nabla_\alpha\omega\biggl(\phi^\mu\frac{dx^\alpha}{d\tau}\biggr).
\end{equation}

The action for the field $B_{\mu\nu}$ is of the Maxwell-Proca form
for a massive vector field $\phi_\mu$. It can be proved that this
theory possesses a stable vacuum and the Hamiltonian is bounded
from below. Even though the action is not gauge invariant, it can
be shown that the longitudinal mode $\phi_0$ (where
$\phi_\mu=(\phi_0,\phi_i)\, (i=1,2,3)$) does not propagate and the
theory is free of ghosts. Similar arguments apply to the MSTG
theory~\cite{Moffat2}.

The Hamilton-Dirac (HD) method is a tool for investigating the
constraints and the degrees of freedom of a field
theory~\cite{Dirac}. The HD procedure checks the theory for
consistency by producing the explicit constraints, and counting
the number of degrees of freedom. It is a canonical initial value
analysis. When the field theory is coupled dynamically to gravity,
many vector field theories are ruled out, because the constraints
produced by the canonical, Cauchy initial value formalism yield
``derivative coupled'' theories (the Christoffel connections do
not cancel out). The Maxwell and Maxwell-Proca theories are prime
examples~\cite{Nester} of consistent vector field theories. With
or without the gravitational field coupling, Maxwell's theory has
two degrees of freedom and Maxwell-Proca has three, and they are
stable and satisfy a consistent Cauchy evolution analysis. Many
other vector field theories are derivative coupled when the
gravitational field is introduced into the action. Severe
singularity problems can appear in vector-gravity coupled
theories, which render them inconsistent. There are no
pathological singularities in the Maxwell-Proca theory coupled to
gravity, when one solves for the second time derivative in the
canonical initial value formulation. In other vector theories,
singularities occur that spoil the stability of the theory and
rule them out as physically unviable theories.

It is possible to attribute the mass $\mu$ to a spontaneous
symmetry breaking mechanism, but we shall not pursue this
possibility at present.

\section{Equations of Motion, Weak Fields and the Modified Gravitational Acceleration}

Let us assume that we are in a distance scale regime for which the
fields $G$, $\omega$ and $\mu$ take their approximate renormalized
constant values:
\begin{equation}
\label{constants} G\sim G_0(1+Z),\quad \omega\sim\omega_0A,\quad
\mu\sim\mu_0B,
\end{equation}
where $G_0, \omega_0$ and $\mu_0$ denote the ``bare'' values of
$G, \omega$ and $\mu$, respectively, and $Z, A$ and $B$ are the
associated renormalization constants.

For a static spherically symmetric field the line element is given
by
\begin{equation}
\label{metric}
ds^2=\gamma(r)dt^2-\alpha(r)dr^2-r^2(d\theta^2+\sin^2\theta
d\phi^2).
\end{equation}
The equations of motion for a test particle obtained from
(\ref{eqsmotion}), (\ref{force}) and (\ref{constants}) are given
by
\begin{equation}
\label{rmotion}
\frac{d^2r}{d\tau^2}+\frac{\alpha'}{2\alpha}\biggl(\frac{dr}{d\tau}\biggr)^2
-\frac{r}{\alpha}
\biggl(\frac{d\theta}{d\tau}\biggr)^2-r\biggl(\frac{\sin^2\theta}{\alpha}\biggr)\biggl(\frac{\sigma
d\phi}{d\tau}\biggr)^2
+\frac{\gamma'}{2\alpha}\biggl(\frac{dt}{d\tau}\biggr)^2
$$ $$
+\sigma\frac{1}{\alpha}\biggl(\frac{d\phi_0}{dr}\biggr)\biggl(\frac{dt}{d\tau}\biggr)=0,
\end{equation}
\begin{equation}
\label{tequation}
\frac{d^2t}{d\tau^2}+\frac{\gamma'}{\gamma}\biggl(\frac{dt}{d\tau}\biggr)
\biggl(\frac{dr}{d\tau}\biggr)
+\sigma\frac{1}{\gamma}\biggl(\frac{d\phi_0}{dr}\biggr)\biggl(\frac{dr}{d\tau}\biggr)=0,
\end{equation}
\begin{equation}
\frac{d^2\theta}{d\tau^2}+\frac{2}{r}\biggl(\frac{d\theta}{d\tau}
\biggr)\biggl(\frac{dr}{d\tau}\biggr)-\sin\theta\cos\theta\biggl(\frac{d\phi}{d\tau}\biggr)^2
=0,
\end{equation}
\begin{equation}
\label{phiequation}
\frac{d^2\phi}{d\tau^2}+\frac{2}{r}\biggl(\frac{d\phi}{d\tau}\biggr)\biggl(\frac{dr}{d\tau}\biggr)
+2\cot\theta\biggl(\frac{d\phi}{d\tau}\biggr)\biggl(\frac{d\theta}{d\tau}\biggr)=0,
\end{equation}
where $\sigma=\lambda\omega/m$.

The orbit of the test particle can be shown to lie in a plane and
by an appropriate choice of axes, we can make $\theta=\pi/2$.
Integrating Eq.(\ref{phiequation}) gives
\begin{equation}
\label{angular} r^2\frac{d\phi}{d\tau}=J,
\end{equation}
where $J$ is the conserved orbital angular momentum. Integration
of Eq.(\ref{tequation}) gives
\begin{equation}
\label{dtequation}
\frac{dt}{d\tau}=-\frac{1}{\gamma}(\sigma\phi_0+E),
\end{equation}
where $E$ is the constant energy per unit mass.

By substituting (\ref{dtequation}) into (\ref{rmotion}) and using
(\ref{angular}), we obtain
\begin{equation}
\label{reducedrmotion}
\frac{d^2r}{d\tau^2}+\frac{\alpha'}{2\alpha}\biggl(\frac{dr}{d\tau}\biggr)^2
-\frac{J^2}{\alpha
r^3}+\frac{\gamma'}{2\alpha\gamma^2}(\sigma\phi_0+E)^2
=\sigma\frac{1}{\alpha\gamma}\biggl(\frac{d\phi_0}{dr}\biggr)(\sigma\phi_0+E).
\end{equation}

We do not have an exact, spherically symmetric static solution to
our field equations for a non-zero $V(\phi)$. However, if we
neglect $V(\phi)$ in Eq.(\ref{phifieldequation}) and $\Lambda$ in
Eq.(\ref{Einsteineqs}), then the exact static, spherically
symmetric (Reissner-Nordstr\"{o}m) solution in empty space yields
the line element
\begin{equation}
\label{RNsolution}
ds^2=\biggl(1-\frac{2GM}{r}+\frac{Q^2}{r^2}\biggr)dt^2-\biggl(1-\frac{2GM}{r}
+\frac{Q^2}{r^2}\biggr)^{-1}dr^2-r^2(d\theta^2+\sin^2\theta
d\phi^2),
\end{equation}
where $M$ is a constant of integration and
\begin{equation}
Q^2=4\pi G\omega\epsilon^2.
\end{equation}
Here, $\epsilon$ denotes the ``charge'' of the spin~-1 vector
particle given by
\begin{equation}
\epsilon=\int d^3xq,
\end{equation}
where the matter current density $J^\mu$ is identified as
$J^\mu=(q,J^i)$.

For large enough values of $r$, the solution (\ref{RNsolution})
approximates the Schwarzschild metric components $\alpha$ and
$\gamma$:
\begin{equation}
\label{Schwarzschild} \alpha(r)\sim \frac{1}{1-2GM/r},\quad
\gamma(r)\sim 1-\frac{2GM}{r}.
\end{equation}
It is not unreasonable to expect that a static, spherically
symmetric solution of the field equations including the mass term
$\mu$ in the field equations (\ref{phifieldequation}) will
approximate for large values of $r$ the Schwarzschild metric
components (\ref{Schwarzschild}).

We assume that $2GM/r\ll 1$ and the slow motion approximation
$dr/ds\sim dr/dt\ll 1$. Then for material test particles, we
obtain from (\ref{eqsmotion}), (\ref{force}), (\ref{constants}),
(\ref{reducedrmotion}) and (\ref{Schwarzschild}):
\begin{equation}
\label{particlemotion}
\frac{d^2r}{dt^2}-\frac{J^2_N}{r^3}+\frac{GM}{r^2}=\sigma\frac{d\phi_0}{dr},
\end{equation}
where $J_N$ is the Newtonian orbital angular momentum.

For weak gravitational fields to first order, the static equations
for $\phi_0$ obtained from (\ref{phiboxequation}) are given for
the source-free case by
\begin{equation}
{\vec\nabla}^2\phi_0-\mu^2\phi_0=0,
\end{equation}
where ${\vec\nabla}^2$ is the Laplacian operator, and we have
neglected any contribution from the self-interaction potential
$W(\phi)$. For a spherically symmetric static field $\phi_0$, we
obtain
\begin{equation}
\phi_0''+\frac{2}{r}\phi_0'-\mu^2\phi_0=0.
\end{equation}
This has the Yukawa solution
\begin{equation}
\phi_0(r)=-\beta\frac{\exp(-\mu r)}{r},
\end{equation}
where $\beta$ is a constant. We obtain from
(\ref{particlemotion}):
\begin{equation}
\label{particlemotion2}
\frac{d^2r}{dt^2}-\frac{J^2_N}{r^3}+\frac{GM}{r^2}=K\frac{\exp(-\mu
r)}{r^2}(1+\mu r),
\end{equation}
where $K=\sigma\beta$.

We observe that the additional Yukawa force term in
Eq.(\ref{particlemotion2}) is {\it repulsive} in accordance with
the exchange of a spin 1 massive boson. We shall find that this
repulsive component of the gravitational field is necessary to
obtain a fit to galaxy rotation curves.

We shall write for the radial acceleration derived from
(\ref{particlemotion2}):
\begin{equation}
a(r)=-\frac{G_{\infty}M}{r^2}+K\frac{\exp(-r/r_0)}{r^2}\biggl(1+\frac{r}{r_0}\biggr),
\end{equation}
and $G_{\infty}$ is defined to be the effective gravitational
constant at infinity
\begin{equation}
\label{renormG}
G_{\infty}=G_0\biggl(1+\sqrt{\frac{M_0}{M}}\biggr).
\end{equation}
Here, $M_0$ denotes a parameter that vanishes when $\omega=0$ and
$G_0$ is Newton's gravitational ``bare'' constant. The constant
$K$ is chosen to be
\begin{equation}
\label{Kequation} K=G_0\sqrt{MM_0}.
\end{equation}
The choice of $K$, which determines the strength of the coupling
of $B_{\mu\nu}$ to matter and the magnitude of the Yukawa force
modification of weak Newtonian gravity, is based on phenomenology
and is not at present derivable from the STVG action formalism.

By using (\ref{renormG}), we can rewrite the acceleration in the
form
\begin{equation}
\label{accelerationlaw} a(r)=-\frac{G_0
M}{r^2}\biggl\{1+\sqrt{\frac{M_0}{M}}\biggl[1-\exp(-r/r_0)
\biggl(1+\frac{r}{r_0}\biggr)\biggr]\biggr\}.
\end{equation}
We can generalize this to the case of a mass distribution by
replacing the factor $G_0M/r^2$ in (\ref{accelerationlaw}) by
$G_0{\cal M}(r)/r^2$. The rotational velocity of a star $v_c$ is
obtained from $v_c^2(r)/r=a(r)$ and is given by
\begin{equation}
v_c=\sqrt{\frac{G_0{\cal
M}(r)}{r}}\biggl\{1+\sqrt{\frac{M_0}{M}}\biggl[1-\exp(-r/r_0)
\biggl(1+\frac{r}{r_0}\biggr)\biggr]\biggr\}^{1/2}.
\end{equation}
The gravitational potential for a point source obtained from the
modified acceleration law (\ref{accelerationlaw}) is given by
\begin{equation}
\Phi(r)=\frac{G_0M}{r}\biggl[1+\sqrt{\frac{M_0}{M}}(1-\exp(-r/r_0))\biggr].
\end{equation}

The acceleration law (\ref{accelerationlaw}) can be written as
\begin{equation}
\label{accelerationGrun} a(r)=-\frac{G(r)M}{r^2},
\end{equation}
where
\begin{equation}
\label{runningG}
G(r)=G_0\biggl[1+\sqrt{\frac{M_0}{M}}\biggl(1-\exp(-r/r_0)
\biggl(1+\frac{r}{r_0}\biggr)\biggr)\biggr]
\end{equation}
is an {\it effective} expression for the variation of $G$ with
respect to $r$. A good fit to a large number of galaxies has been
achieved with the parameters~\cite{Moffat2,Brownstein}:
\begin{equation}
M_0=9.60\times 10^{11}\,M_{\odot},\quad r_0=13.92\,{\rm
kpc}=4.30\times 10^{22}\,{\rm cm}.
\end{equation}
In the fitting of the galaxy rotation curves for both LSB and HSB
galaxies, using photometric data to determine the mass
distribution ${\cal M}(r)$~\cite{Brownstein}, only the
mass-to-light ratio $\langle M/L\rangle$ is employed, once the
values of $M_0$ and $r_0$ are fixed universally for all LSB and
HSB galaxies. Dwarf galaxies are also fitted with the
parameters~\cite{Brownstein}:
\begin{equation}
M_0=2.40\times 10^{11}\,M_{\odot},\quad r_0=6.96\,{\rm
kpc}=2.15\times 10^{22}\,{\rm cm}.
\end{equation}
By choosing values for the parameters $G_{\infty}$, $(M_0)_{\rm
clust}$ and $(r_0)_{\rm clust}$, we are able to obtain
satisfactory fits to a large sample of X-ray cluster
data~\cite{Brownstein2}.

\section{Orbital Equations of Motion}

We set $\theta=\pi/2$ in (\ref{metric}), divide the resulting
expression by $d\tau^2$ and use Eqs.(\ref{angular}) and
(\ref{dtequation}) to obtain
\begin{equation}
\label{energyconserved}
\biggl(\frac{dr}{d\tau}\biggr)^2+\frac{J^2}{\alpha
r^2}-\frac{1}{\alpha\gamma}(\sigma\phi_0+E)^2=-\frac{E}{\alpha}.
\end{equation}
We have $ds^2=Ed\tau^2$, so that $ds/d\tau$ is a constant. For
material particles $E>0$ and for massless photons $E=0$.

Let us set $u=1/r$ and by using (\ref{angular}), we have
$dr/d\tau=-Jdu/d\phi$. Substituting this into
(\ref{energyconserved}), we obtain
\begin{equation}
\label{neworbital} \biggl(\frac{du}{d\phi}\biggr)^2=
\frac{1}{\alpha\gamma J^2}(E+\sigma\phi_0)^2-\frac{1}{\alpha
r^2}-\frac{E}{\alpha J^2}.
\end{equation}
On substituting (\ref{Schwarzschild}) and
$dr/d\phi=-(1/u^2)du/d\phi$ into (\ref{neworbital}), we get after
some manipulation:
\begin{equation}
\label{finalorbital}
\frac{d^2u}{d\phi^2}+u=\frac{EGM}{J^2}-\frac{EK}{J^2}
\exp\biggl(-\frac{1}{r_0u}\biggr)\biggl(1+\frac{1}{r_0u}\biggr)
+3GMu^2,
\end{equation}
where $r_0=1/\mu$.

For material test particles $E=1$ and we obtain
\begin{equation}
\label{materialorbit}
\frac{d^2u}{d\phi^2}+u=\frac{GM}{J^2}+3GMu^2-\frac{K}{J^2}\exp\biggl(-\frac{1}{r_0u}\biggr)
\biggl(1+\frac{1}{r_0u}\biggr).
\end{equation}
On the other hand, for massless photons $ds^2=0$ and $E=0$ and
(\ref{finalorbital}) gives
\begin{equation}
\label{photons} \frac{d^2u}{d\phi^2}+u=3GMu^2.
\end{equation}

\section{Solar System and Binary Pulsar Observations}

We obtain from Eq.(\ref{materialorbit}) the orbit equation (we
reinsert the speed of light c):
\begin{equation}
\label{particleorbit} \frac{d^2u}{d\phi^2}+u=\frac{GM}{c^2
J^2}-\frac{K}{c^2J^2}\exp(-r/r_0)\biggl[1
+\biggl(\frac{r}{r_0}\biggr)\biggr]+\frac{3GM}{c^2}u^2.
\end{equation}
Using the large $r$ weak field approximation, and the expansion
\begin{equation}
\exp(-r/r_0)=
1-\frac{r}{r_0}+\frac{1}{2}\biggl(\frac{r}{r_0}\biggr)^2+...
\end{equation}
we obtain the orbit equation for $r\ll r_0$:
\begin{equation}
\label{orbitperihelion}
\frac{d^2u}{d\phi^2}+u=N+3\frac{GM}{c^2}u^2,
\end{equation}
where
\begin{equation}
N=\frac{GM}{c^2J_N^2}-\frac{K}{c^2J_N^2}.
\end{equation}

We can solve Eq.(\ref{orbitperihelion}) by perturbation theory and
find for the perihelion advance of a planetary orbit
\begin{equation}
\label{perihelionformula} \Delta\omega=\frac{6\pi}{c^2L}
(GM_{\odot}-K_{\odot}),
\end{equation}
where $J_N=(GM_{\odot}L/c^2)^{1/2}$, $L=a(1-e^2)$ and $a$ and $e$
denote the semimajor axis and the eccentricity of the planetary
orbit, respectively.

We now use the running of the effective gravitational coupling
constant $G=G(r)$, determined by (\ref{runningG}) and find that
for the solar system $r\ll r_0$, we have $G\sim G_0$ within the
experimental errors for the measurement of Newton's constant
$G_0$. We choose for the solar system
\begin{equation}
\label{perhbound} \frac{K_{\odot}}{c^2}\ll 1.5\,{\rm km}
\end{equation}
and use $G=G_0$ to obtain from (\ref{perihelionformula}) a
perihelion advance of Mercury in agreement with GR. The bound
(\ref{perhbound}) requires that the coupling constant $\omega$
varies with distance in such a way that it is sufficiently small
in the solar system regime and determines a value for $M_0$, in
Eq.(\ref{Kequation}), that is in accord with the bound
(\ref{perhbound}).

For terrestrial experiments and orbits of satellites, we have also
that $G\sim G_0$ and for $K_{\oplus}$ sufficiently small, we then
achieve agreement with all gravitational terrestrial experiments
including E\"otv\"os free-fall experiments and ``fifth force''
experiments.

For the binary pulsar PSR 1913+16 the formula
(\ref{perihelionformula}) can be adapted to the periastron shift
of a binary system. Combining this with the STVG gravitational
wave radiation formula, which will approximate closely the GR
formula, we can obtain agreement with the observations for the
binary pulsar.  The mean orbital radius for the binary pulsar is
equal to the projected semi-major axis of the binary, $\langle
r\rangle_N=7\times 10^{10}\,{\rm cm}$, and we choose $\langle
r\rangle_N\ll r_0$. Thus, for $G=G_0$ within the experimental
errors, we obtain agreement with the binary pulsar data for the
periastron shift when
\begin{equation}
\label{binarybound} \frac{K_N}{c^2}\ll 4.2\,{\rm km}.
\end{equation}

For a massless photon $E=0$ and we have
\begin{equation}
\label{lightbending} \frac{d^2u}{d\phi^2}+u=3\frac{GM}{c^2}u^2.
\end{equation}
For the solar system using $G=G_0$ within the experimental errors
gives the light deflection:
\begin{equation}
\Delta_{\odot}=\frac{4G_0M_{\odot}}{c^2R_{\odot}}
\end{equation}
in agreement with GR.

\section{Galaxy Clusters and Lensing}

The bending angle of a light ray as it passes near a massive
system along an approximately straight path is given to lowest
order in $v^2/c^2$ by
\begin{equation}
\label{lensingformula} \theta=\frac{2}{c^2}\int\vert
a^{\perp}\vert dz,
\end{equation}
where $\perp$ denotes the perpendicular component to the ray's
direction, and dz is the element of length along the ray and $a$
denotes the acceleration.

From (\ref{lightbending}), we obtain the light deflection
\begin{equation}
\Delta=\frac{4GM}{c^2R}=\frac{4G_0{\overline M}}{c^2R},
\end{equation}
where
\begin{equation}
{\overline M}=M\biggl(1+\sqrt{\frac{M_0}{M}}\biggr).
\end{equation}
The value of ${\overline M}$ follows from (\ref{runningG}) for
clusters as $r\gg r_0$ and
\begin{equation}
G(r)\rightarrow
G_{\infty}=G_0\biggl(1+\sqrt{\frac{M_0}{M}}\biggr).
\end{equation}
We choose for a cluster $M_0=3.6\times 10^{15}\,M_{\odot}$ and a
cluster mass $M_{\rm clust}\sim 10^{14}\,M_{\odot}$, and obtain
\begin{equation}
\biggl(\sqrt{\frac{M_0}{M}}\biggr)_{\rm clust}\sim 6.
\end{equation}
We see that ${\overline M}\sim 7M$ and we can explain the increase
in the light bending without exotic dark matter.

From the formula Eq.(\ref{accelerationlaw}) for $r\gg r_0$ we get
\begin{equation}
a(r)=-\frac{G_0\overline M}{r^2}.
\end{equation}
We expect to obtain from this result a satisfactory description of
lensing phenomena using Eq.(\ref{lensingformula}).

An analysis of a large number of clusters shows that the MSTG and
STVG theories fit well the cluster data in terms of the cluster
mass, $M_{\rm clust}$, and an average value for the parameter
$M_0$~\cite{Brownstein2}.

\section{Running of the Effective Constants $G$, $\omega$ and $\mu$}

The scaling with distance of the effective gravitational constant
$G$, the effective coupling constant $\omega$ and the effective
mass $\mu$ is seen to play an important role in describing
consistently the solar system and the galaxy and cluster dynamics,
without the postulate of exotic dark matter. We have to solve the
field equations (\ref{Gequation}), (\ref{omegaequation}) and
(\ref{muequation}) with given potentials $V(G), V(\omega)$ and
$V(\mu)$ to determine the variation of the effective constants
with space and time. These equations are complicated, so we shall
make simplifying approximations. In Eq.(\ref{Gequation}), we shall
neglect the contributions from $N$ and obtain for $T=\Lambda=0$:
\begin{equation}
\label{Gequation2} \nabla_\nu\nabla^\nu G+V'(G)=0,
\end{equation}
where $f'(y)=df/dy$. The effective variation of $G$ with $r$ is
determined by Eq.(\ref{runningG}). We obtain from
(\ref{Gequation2}) for the static spherically symmetric equations
\begin{equation}
\label{nabla2G} {\vec\nabla}^2G(r) -V'(G)\equiv
G''(r)+\frac{2}{r}G'(r)-V'(G)=0.
\end{equation}
By choosing the potential
\begin{equation}
V(G)=-\frac{1}{2}\biggl(\frac{G_0}{r_0^2}\biggr)\biggl(\frac{M_0}{M}\biggr)\exp(-2r/r_0)
\biggr(1+\frac{2r}{r_0}-\frac{r^2}{r_0^2}\biggr),
\end{equation}
we obtain a solution to (\ref{nabla2G}) for $G(r)$ given by
(\ref{runningG}). The neglect of the contributions $N$ can only be
justified by solving the complete set of coupled equations by a
perturbation calculation. We will not attempt to do this in the
present work, but we plan to investigate this issue in a future
publication.

We see from (\ref{runningG}) that for $r\ll r_0$ we obtain
$G(r)\sim G_0$. As the distance scale approaches the regime of the
solar system $r < 100 A.U.$ where $1 A.U. = 1.496\times
10^{13}\,{\rm cm}=4.85\times 10^{-9}\,{\rm kpc}$, then
(\ref{accelerationlaw}) becomes the Newtonian acceleration law:
\begin{equation}
\label{NewtonG} a(r)=-\frac{G_0M}{r^2},
\end{equation}
in agreement with solar physics observations for the inner
planets.

Let us make the approximation of neglecting $F(\phi)$ in
Eq.(\ref{omegaequation}). In the static spherically symmetric case
this gives
\begin{equation}
\label{nabla2mu} \omega''(r)+\frac{2}{r}\omega'(r)-V'(\omega)=0.
\end{equation}
We choose as a solution for $\omega(r)$:
\begin{equation}
\label{omegasolution}
\omega(r)=\omega_0\{1+{\overline\omega}[1-\exp(-{\overline\mu r})
(1+{\overline\mu}r)]\},
\end{equation}
where $\overline\omega$ and $\overline\mu$ are positive constants.
The potential $V(\omega)$ has the form
\begin{equation}
V(\omega)=-\frac{1}{2}\omega_0^2{\overline\mu}^2{\overline\omega}^2\exp(-2\overline\mu
r)(1+2\overline\mu r-{\overline\mu}^2r^2).
\end{equation}

For the variation of the renormalized mass $\mu=\mu(r)$, we find
that a satisfactory solution to Eq.(\ref{muequation}) should
correspond to a a $\mu(r)=1/r_0(r)$ that decreases from a value
for the inner planets of the solar system, consistent with solar
system observations, to a small value corresponding to $r_0$ for
the galaxy fits, $r_0=14$ kpc, and to an even smaller value for
the cluster data fits.

The spatial variations of $G(r)$, $\omega(r)$ and
$\mu(r)=1/{r_0(r)}$ can be determined numerically from the
equations (\ref{Gequation}), (\ref{omegaequation}) and
(\ref{muequation}) with given potentials $V(G), V(\omega)$ and
$V(\mu)$, such that for the solar system and the binary pulsar PSR
1913+16 the bounds (\ref{perhbound}) and (\ref{binarybound}) are
satisfied by the solutions for $G(r), \omega(r)$ and $\mu(r)$. The
spatial variations of $G$, the coupling constant $\omega$ and the
range $r_0$ are required to guarantee consistency with solar
system observations. On the other hand, their increase at galactic
and cosmological distance and time scales can account for galaxy
rotation curves, cluster lensing and cosmology without
non-baryonic dark matter.

We have constructed a classical action for gravity that can be
considered as an {\it effective} field theory description of an RG
flow quantum gravity scenario as described in refs.~\cite{Reuter}
and~\cite{Moffat2}.

The fitting of the solar system, galaxy and the clusters of
galaxies data depends on the running of the of the ``constants''
$G$, $\omega$ and $r_0$. They should increase from one distance
scale to the next according to the renormalization group flow
diagrams, or the solutions of the classical field equations in the
present article. In a future article, the author plans to provide
a more complete determination of the running of the constants.
However, the present article describes the basic scenario and the
ideas underlying the theory.

\section{Cosmology}

Let us now consider a cosmological solution to our STVG theory. We
adopt a homogeneous and isotropic FLRW background geometry with
the line element
\begin{equation}
ds^2=dt^2-a^2(t)\biggl(\frac{dr^2}{1-kr^2}+r^2d\Omega^2\biggr),
\end{equation}
where $d\Omega^2=d\theta^2+\sin^2\theta d\phi^2$ and $k=0,-1,+1$
for a spatially flat, open and closed universe, respectively. In
this background spacetime, we have $\phi_0\equiv\psi\not= 0$,
$\phi_i=0$ and $B_{\mu\nu}=0$.

We define the energy-momentum tensor for a perfect fluid by
\begin{equation}
T^{\mu\nu}=(\rho+p)u^\mu u^\nu-pg^ {\mu\nu},
\end{equation}
where $u^\mu=dx^\mu/ds$ is the 4-velocity of a fluid element and
$g_{\mu\nu}u^\mu u^\nu=1$. Moreover, we have
\begin{equation}
\rho=\rho_M+\rho_\phi+\rho_S,\quad p=p_M+p_\phi+p_S,
\end{equation}
where $\rho_i$ and $p_i$ denote the components of density and
pressure associated with the matter, the $\phi^\mu$ field and the
scalar fields $G$, $\omega$ and $\mu$, respectively.

The Friedmann equations take the form
\begin{equation}
H^2(t)+\frac{k}{a^2(t)}=\frac{8\pi G(t)\rho(t)}{3}+\frac{{\dot
a}}{a}\frac{{\dot G}}{G}+\frac{\Lambda}{3},
\end{equation}
\begin{equation}
\frac{{\ddot a}(t)}{a(t)}=-\frac{4\pi
G(t)}{3}(\rho(t)+3p(t))+\frac{1}{2}\biggl(\frac{{\ddot
G}}{G}-\frac{{\dot G}^2}{G^2}+\frac{2{\dot a}}{a}\frac{{\dot
G}}{G}\biggr) +\frac{\Lambda}{3},
\end{equation}
where $H(t)={\dot a}(t)/a(t)$.

Let us make the simplifying approximation for equations
(\ref{Gequation}):
\begin{equation}
\label{approxGeq} {\ddot{\cal G}}+3H{\dot{\cal G}}+V'({\cal
G})=\frac{1}{2}G_0{\cal G}^2\biggl[\rho-3p+\frac{\Lambda}{4\pi
G_0{\cal G}}\biggr],
\end{equation}
where ${\cal G}(t)=G(t)/G_0$. A solution for ${\cal G}$ in terms
of a given potential $V({\cal G})$ and for given values of $\rho$
and $p$ can be obtained from (\ref{approxGeq}).

The solution for ${\cal G}$ must satisfy a constraint at the time
of big bang nucleosynthesis~\cite{Bean}. The number of
relativistic degrees of freedom is very sensitive to the cosmic
expansion rate at 1 MeV. This can be used to constrain the time
dependence of $G$. Recent measurements of the $^4He$ mass fraction
and the deuterium abundance at 1 MeV leads to the constraint
$G(t)\sim G_0$. We impose the conditions ${\cal G}(t)\rightarrow
1$ as $t\rightarrow t_{BBN}$ and ${\cal G}(t)\rightarrow 1+{\bar
\omega}$ as $t\rightarrow t_{SLS}$ where $t_{BBN}$ and $t_{SLS}$
denote the times of the big bang nucleosynthesis and the surface
of last scattering, respectively. A possible solution for ${\cal
G}$ can take the form
\begin{equation}
\label{Gsolution} {\cal G}(t)
=1+{\bar\omega}\biggl[1-\exp(-t/T)\biggl(1+\frac{t}{T}\biggr)\biggr],
\end{equation}
where ${\bar\omega}$ and $T$ are constants and ${\bar\omega}$ is a
measure of the magnitude of the scalar field $\psi$. We have for
$t >> T$ that ${\cal G}\rightarrow 1+{\bar\omega}$ and for $t <<
T$ that ${\cal G}\rightarrow 1$. We get
\begin{equation}
{\dot{\cal G}}=\frac{\bar\omega t}{T^2}\exp(-t/T),\quad
{\ddot{\cal
G}}=\frac{{\bar\omega}}{T^2}\exp(-t/T)\biggl(1-\frac{t}{T}\biggr).
\end{equation}
It follows that ${\dot{\cal G}}(t)\rightarrow 0$ for $t >> T$,
which allows us for a suitable choice of $T$ to satisfy the
experimental bound from the Cassini spacecraft
measurements~\cite{Cassini}:
\begin{equation}
\vert{\dot G}/G\vert \simeq 10^{-13}\,{\rm yr}^{-1}.
\end{equation}

A linear perturbation on the FLRW background will link the theory
with observations of anisotropies in the CMB as well as galaxy
clustering on large scales. The basic fields are perturbed around
the background spacetime (denoted for a quantity $Y$ by ${\tilde
Y}$). In the conformal metric with the time transformation
$d\eta=dt/a(t)$:
\begin{equation}
ds^2=a^2(\eta)(d\eta^2-d{\vec x}^2),
\end{equation}
the metric perturbations are in the conformal Newtonian gauge
\begin{equation}
g_{00}({\vec x},t)=a^2(t)(1+2\Phi({\vec x},t)),\quad g_{ij}({\vec
x},t)=a^2(t)(1-2\Phi({\vec x},t))\delta_{ij},
\end{equation}
where $\Phi$ is the gravitational potential. The vector field
perturbations are defined by
\begin{equation}
\phi_\mu({\vec x},t)=a(t)({\tilde\phi}_\mu
(t)+\delta\phi_\mu({\vec x},t)),
\end{equation}
where ${\tilde\phi}_i(t)=0$. Denoting  by $\chi_i$ the scalar
fields $\chi_1=G$, $\chi_2=\omega$ and $\chi_3=\mu$, the scalar
field perturbations are
\begin{equation}
\chi_i({\vec x},t)={\tilde\chi}_i(t)+\delta\chi_i({\vec x},t).
\end{equation}

The problem that gravity theories such as MSTG and STVG face in
describing cosmology with no cold dark matter (CDM) (non-baryonic
dark matter) is the damping of perturbations during the
recombination era. In a pure baryonic universe evolving according
to Einstein's gravitational field equations, the coupling of
baryons to photons during the recombination era will suffer Silk
damping, causing the collisional propagation of radiation from
overdense to underdense regions~\cite{Silk,Skordis}. In the CDM
model, the perturbations $\delta_{CDM}$ are undamped during
recombination, because the CDM particles interact with gravity and
only weakly with matter (photons). The Newtonian potential in the
CDM model is approximately given by
\begin{equation}
k^2\Phi\sim 4\pi G_0(\rho_b\delta_b+\rho_{CDM}\delta_{CDM}),
\end{equation}
where $k^2$ denotes the square of the wave number, $\rho_b$ and
$\rho_{CDM}$ denote the densities of baryons and cold dark matter,
respectively, and $\delta_i$ denotes the perturbation density
contrast for each component i of matter. If $\rho_{CDM}$ is
sufficiently large, then $\delta_{CDM}$ will not be erased,
whereas $\delta_b$ decreases during recombination~\cite{Peebles2}.
In STVG, the imperfect fluid plasma before recombination has two
components: the dominant neutral vector field component that does
not couple to photons and the baryon-photon component. The vector
field component has zero pressure and zero shear viscosity, so the
vector field perturbations are not Silk damped like the
baryon-radiation perturbations, for the latter have non-vanishing
pressure and shear viscosity.

In STVG, we have
\begin{equation}
\Omega_b=\frac{8\pi G\rho_b}{3H^2},\quad \Omega_\psi =\frac{8\pi
G\rho_\psi}{3H^2},\quad \Omega_S=\frac{8\pi G\rho_S}{3H^2}.
\end{equation}
We also have a possible contribution from massive neutrinos
\begin{equation}
\Omega_\nu=\frac{8\pi G\rho_\nu}{3H^2},
\end{equation}
where $\rho_\nu$ denotes the density of neutrinos. We assume that
the vector field density dominates.

The Newtonian potential in our modified gravitational model
becomes
\begin{equation}
\label{STVGpert} k^2\Phi\sim 4\pi G_{\rm
ren}[\rho_b\delta_b+\rho_\nu\delta_\nu
+\rho_\psi\delta_\psi+\rho_S\delta_S],
\end{equation}
where $G_{\rm ren}$ is the renormalized value of the gravitational
constant. Assuming that the density $\rho_\psi$ is significant
before and during recombination, we can consider fitting the
acoustic peaks in the power spectrum in a spatially flat universe
with the parameters
\begin{equation}
\Omega=\Omega_b+\Omega_\nu+\Omega_\psi+\Omega_\nu
+\Omega_\Lambda=1.
\end{equation}

The fitting of the acoustic peaks in the CMB power spectrum does
not permit a too large value of $\Omega_\Lambda$. Moreover, the
neutrino contribution is constrained by the three neutrinos having
a mass $ < 2$ eV. We could choose, for example, $\Omega_b=0.04,
\Omega_\psi=0.25,\Omega_\nu =0.01, \Omega_\Lambda=0.7$ as a
possible choice of parameters to fit the data. There are now new
data from the balloon borne Boomerang CMB
observations~\cite{balloon} that together with other ground based
observations and WMAP data determine more accurately the height of
the third acoustic peak in the angular CMB power
spectrum~\cite{Silk2}. The ratio of the height of the first peak
to the second peak determines the baryon content $\Omega_b\sim
0.04$. The height of the third peak is determined by the amount of
cold dark matter, and in the modified gravity theory by the
possible amounts of scalar $\psi$, $G$, $\omega$ and $\mu$
contributions. In particular, the dominant neutral $\psi$ vector
component perturbations will not be washed out before
recombination.

Can the effects of gravitational constant renormalization together
with the possible effects of the densities $\rho_\psi$ and
$\rho_S$ describe a universe which can reproduce the current
galaxy and CMB observations?  The answer to this problem will be
addressed in a future publication.

\section{Conclusions}

A modified gravity theory based an a $D=4$ pseudo-Riemannian
metric, a spin $1$ vector field and a corresponding second-rank
skew field $B_{\mu\nu}$ and dynamical scalar fields $G$, $\omega$
and $\mu$, yields a static spherically symmetric gravitational
field with an added Yukawa potential and with an effective
coupling strength and distance range. This modified acceleration
law leads to remarkably good fits to a large number of
galaxies~\cite{Brownstein} and galaxy clusters~\cite{Brownstein2}.
The previously published gravitational theories NGT~\cite{Moffat}
and MSTG~\cite{Moffat2} yielded the same modified weak
gravitational field acceleration law and, therefore, the same
successful fits to galaxy and cluster data. The MSTG and STVG
gravity theories can both be identified generically as
metric-skew-tensor gravity theories, for they both describe
gravity as a metric theory with an additional degree of freedom
associated with a skew field coupling to matter. For MSTG, this is
a third-rank skew field $F_{\mu\nu\lambda}$, while for STVG the
skew field is a second-rank tensor $B_{\mu\nu}$. However, MSTG is
distinguished from STVG as being the weak field approximation to
the nonsymmetric gravitational theory (NGT).

An action $S_S$ for the scalar fields $G(x)$, $\omega(x)$ and
$\mu(x)=1/r_0(x)$ and the field equations resulting from a
variation of the action, $\delta S_S=0$, can be incorporated into
the NGT and MSTG theory. The dynamical solutions for the scalar
fields give an {\it effective} description of the running of the
constants in an RG flow quantum gravity scenario, in which strong
infrared renormalization effects and increasing large scale
spatial values of $G$ and $\omega$ lead, together with the
modified acceleration law, to a satisfactory description of galaxy
rotation curves and cluster dynamics without non-baryonic dark
matter.

We have demonstrated that a cosmological model with the
renormalized gravitational constant $G_{\rm ren}$ and
contributions from the scalar fields $G(t)$, $\omega(t)$ and
$\mu(t)$ can possibly lead to a satisfactory description of the
distribution of galaxies and the CMB power spectrum in a baryon
dominated universe.

The neutral vector particle $\phi$ does not couple to radiation
and it has zero pressure $p$ and zero shear viscosity. Since it
dominates the period of recombination, its perturbations
associated with the plasma fluid will not be washed out by Silk
damping. In contrast to standard dark matter models, we should not
search for {\it new stable particles} such as weakly interacting
massive particles (WIMPS) or neutralinos, because the fifth force
charge in STVG that is the source of the neutral vector field
(skew field) is carried by the known stable baryons (and electrons
and neutrinos). This new charge is the source of a fifth force
skew field that modifies the gravitational field in the universe.

\vskip 0.2 true in {\bf Acknowledgments} \vskip 0.2 true in

This work was supported by the Natural Sciences and Engineering
Research Council of Canada. I thank Yujun Chen, Joel Brownstein
Martin Green and Pierre Savaria for helpful discussions.

\end{document}